\documentclass[12pt,letterpaper]{article}         
\usepackage{jheppub}
\usepackage{epsfig}

\def\be{\begin{eqnarray}}
\def\ee{\end{eqnarray}}

\newcommand{\eqn}[1]{(\ref{#1})}

\def\Dslash{\,\,{\raise.15ex\hbox{/}\mkern-12mu D}}
\def\Dbarslash{\,\,{\raise.15ex\hbox{/}\mkern-12mu {\bar D}}}
\def\delslash{\,\,{\raise.15ex\hbox{/}\mkern-9mu \partial}}
\def\delbarslash{\,\,{\raise.15ex\hbox{/}\mkern-9mu {\bar\partial}}}
\def\pslash{\,\,{\raise.15ex\hbox{/}\mkern-9mu p}}
\def\calDslash{\,\,{\raise.15ex\hbox{/}\mkern-12mu {\cal D}}}

\def\lae{\mathrel{\mathop{\smash{\lower .5 ex \hbox{$\stackrel<\sim$}}}}}
\def\lae{\mathrel{\mathop{\smash{\lower .5 ex \hbox{$\stackrel>\sim$}}}}}

\preprint{DCPT-16/49, MIT-CTP/4853}

\title{\large Diffusion and Chaos from near AdS$_2$ horizons}

\author{Mike Blake$^{1}$ and Aristomenis Donos$^{2}$} 

\affiliation{ $^{1}$ Center for Theoretical Physics, Massachusetts Institute of Technology,
Cambridge, MA 02139, USA}

\affiliation{ $^{2}$ Centre for Particle Theory and Department of Mathematical Sciences,
Durham University, Durham, DH1 3LE, U.K.}

\emailAdd{mab90@mit.edu, aristomenis.donos@durham.ac.uk }

\abstract{ We calculate the thermal diffusivity $D = \kappa/c_{\rho}$ and butterfly velocity $v_B$ in holographic models that flow to AdS$_2 \times R^{d}$ fixed points in the infra-red. We show that both these quantities are governed by the same irrelevant deformation of AdS$_2$ and hence establish a simple relationship between them. When this deformation corresponds to a universal dilaton mode of dimension $\Delta = 2$ then this relationship is always given by $D = v_B^2/(2 \pi T)$.}

\begin{document}
\maketitle
\pagestyle{plain} \setcounter{page}{1}
\newcounter{bean}
\baselineskip16pt
\section{Introduction} 
\paragraph{} It has long been suggested that transport in strongly coupled systems is governed by a `smallest possible' relaxation timescale $\tau \sim \hbar/(k_B T)$ \cite{subirbook, sachdev, planckian, hartnoll}. In particular, this timescale has been experimentally detected in both the linear resistivity \cite{mackenzie} and thermal diffusivity \cite{zhang} of strongly correlated materials. Compellingly, the essential logic behind this proposal resonates with the huge recent progress in studying quantum chaos. In that context, one can characterise the growth of the quantum butterfly effect \cite{butterflyeffect, stringyeffects, localised} through a Lyapunov rate $\lambda_L$ for which a rigorous Planckian bound $\lambda_L \leq 2 \pi k_B T/\hbar$ has been formulated\footnote{Henceforth we set $\hbar = k_B = 1$.} \cite{chaos}. 
\paragraph{} Recently it has emerged that in many theories there appear to be further connections between transport properties and chaos \cite{butterfly, incoherent,  aleiner, swingle, patel, stanford}. Specifically in \cite{butterfly, incoherent} it was found that the thermoelectric diffusion constants of many holographic theories are closely related to the butterfly velocity, $v_B$, which describes the speed at which chaos propagates. Subsequently this connection has also been seen in weakly coupled Fermi-liquids \cite{aleiner}, diffusive metals \cite{swingle} and critical Fermi-surface models \cite{patel} and its relevance for understanding the thermal diffusivity of cuprate strange metals was discussed in \cite{zhang}. However it remains unclear how fundamental the connection between chaos and diffusion is, and in particular for what class of theories one might expect to be able to make these observations precise. 
\paragraph{}  Lately there has been an explosion in activity in studying large-$N$ systems, such as Sachdev-Ye-Kitaev (SYK) models or AdS$_2$ holography, which have an approximate $0+1 d$ conformal symmetry \cite{sachdevye, pacoulet, kitaev, sachdevbh, Polchinski2016xgd, Jevicki2016bwu, comments, yang, jensen, stanford, polchinski,Engelsoy:2016xyb,Cvetic:2016eiv}. For such systems, much of the infra-red physics is dominated by the Goldstone mode associated with the fact that the ground state `spontaneously breaks' this symmetry. As such these models provide a simplified context in which one might hope to establish sharp connections between transport properties and chaos\footnote{The connection between diffusion and $v_B$ in a coupled SYK model has recently been studied in \cite{stanford}. We will comment on the relationship to our results in Section~\ref{discussion}.}. 
\paragraph{}  With this in mind, the purpose of this paper is to clarify the precise relationship between diffusion and chaos in holographic theories that approach an AdS$_2 \times R^{d}$ fixed point in the infra-red. In particular we will discuss a general class of models in which we can construct such fixed points either by introducing a finite density for the boundary theory or through `Q-lattice' fields that break translational symmetry \cite{andrade, donos1, donos2, blaise2}. For these theories we can define a thermal diffusivity, $D$, via the Einstein relation
\be
D   = \frac{\kappa}{c_\rho}
\label{thermaldiffusivity}
\ee
where $\kappa$ is the thermal conductivity and $c_{\rho}$ is the specific heat at fixed charge density. Usually this form of the Einstein relation is only valid for particle-hole symmetric theories. However, as we establish in Appendix~\ref{sec:einsteinrelations}, it can also be used to define a thermal diffusivity when our models flow to a finite density AdS$_2\times R^{d}$. 
\paragraph{} Crucially both the diffusion constant \eqn{thermaldiffusivity} and the butterfly velocity $v_B$ are infra-red quantities that can be determined from a near AdS$_2$ horizon. In particular it has been shown in \cite{donos5, donos6} that at any temperature $\kappa$ is generically fixed by the black hole horizon data. However, unlike in the majority of holographic theories studied in \cite{butterfly,incoherent}, we cannot extract $c_{\rho}$ and $v_B$ directly from the fixed point - rather they are controlled by irrelevant deformations to the geometry. A key result of this paper is then that precisely the same irrelevant deformation of AdS$_2 \times R^{d}$ governs the behaviour of both $c_{\rho}$ and $v_{B}$. This allows us establish a simple quantitative relationship
\be
D = E \frac{v_B^2}{2 \pi T}
\label{butterflyresult}
\ee
where $1/2 < E \leq 1$ is a constant that depends only on the scaling dimension of the leading irrelevant deformation.   For generic flows this mode corresponds to a universal bulk dilaton field that parameterises the volume of the black hole horizon. In this case \eqn{butterflyresult} holds with a universal constant $E = 1$. Note that the fact $E$ remains bounded for more general deformations is highly non-trivial. In particular, whilst the infra-red scaling dimension of modes depends on the UV data of the boundary theory, we find that this never significantly changes the relationship between $D$ and $v_B$. 
 \paragraph{} We emphasise that the relationship \eqn{butterflyresult} is valid for a very general class of AdS$_2$ models. In particular, it does not depend on the matter fields that we use to support our geometry. Specifically we will establish that it is true both for translationally invariant theories dual to electric-AdS$_2$ geometries and also when Q-lattice fields are supporting the extremal geometry. Indeed the relationship \eqn{butterflyresult} holds whenever our theories flow to one of these AdS$_2 \times R^{d}$ fixed points.  
\section{AdS$_2 \times R^{d}$ Fixed Points } 
\label{sec:qlattices}
\paragraph{}  As we discussed in the introduction, our goal in this paper is to study the thermal diffusivity \eqn{thermaldiffusivity} and butterfly velocity in a general class of gravitational theories that admit AdS$_2 \times R^{d}$ solutions . In particular we will work with the following action
\begin{eqnarray}\label{eq:Qaction}
S=\int d^{d+2}x\,\sqrt{-g}\,\left(R-\frac{1}{2}(\partial\varphi)^{2}-V(\varphi)-\frac{1}{2}W(\varphi)  (\partial\chi_{\cal A} )^{2}-\frac{1}{4}Z(\varphi)\,F^{2} \right)  
\end{eqnarray}
where the index ${\cal A}$ runs over the $d$ spatial dimensions of the boundary theory. Whilst we will use the specific action \eqn{eq:Qaction} for concreteness, much of our discussion can be straightforwardly generalised to more complicated models such as those with additional scalars. 
\paragraph{} Then the above action admits homogeneous solutions satisfying the ansatz
\begin{eqnarray}\label{eq:Qansatz}
ds^{2}_{d+2} &=&-f(r)\,dt^{2}+\frac{dr^2}{f(r)} +h(r) dx_{\cal A}^2  \nonumber \\
A=a(r) &dt& \;\;\;\;\;\; \chi_{\cal A}= k x_{\cal A} \;\;\;\;\;\;\;\;\; \varphi = \varphi(r) 
\end{eqnarray}
where as usual the constant radial flux of the Maxwell field can be identified with the charge density, $\rho$, of the boundary theory
\be
\rho = Z(\varphi) h^{d/2} a'
\ee
We can also see that when $k \neq 0$ then we have broken the translational symmetry of the boundary theory. Indeed, in this case it is convenient to identify $\varphi$ and $\chi_{\cal A}$ with complex scalar fields $\Psi_{\cal A} = \varphi e^{i \chi_{\cal A}}$ and hence view the above solutions as periodic `Q-lattices' \cite{donos1}. The advantage of breaking this symmetry is that it will enable us to obtain finite expressions for all the thermoelectric response coefficients. However, it is not essential to our analysis and by taking the limit $k \to 0$ we can discuss translationally invariant theories. 
\paragraph{} The main reason we wish to focus on these models is then that the above action admits a very general class of extremal black hole solutions. %
Specifically, the equations of motion admit zero temperature solutions corresponding to an AdS$_2 \times R^{d}$ metric
\begin{align} \label{eq:Ext_Limit}
f=L\,(r-r_{0})^{2},\quad h=h_{0}, \quad \varphi=\varphi_{0} 
\end{align}
provided we satisfy the constraints 
\begin{align}\label{eq:ext_constraints}
0&= - 2 L + \frac{\rho^{2}}{h_{0}^{d} Z(\varphi_{0})}+\frac{k^{2} W(\varphi_{0})}{h_{0}}\notag\\
0&=-2 V(\varphi_{0}) - \frac{\rho^{2}}{h_{0}^{d}Z(\varphi_{0})} -\frac{d k^{2} W(\varphi_{0})}{h_0}\notag\\
0&=-2 V^{\prime}(\varphi_{0})+\frac{ \,Z^{\prime}(\varphi_{0})\ \rho^{2}}{h_{0}^{d}Z(\varphi_{0})^{2}} - \frac{d k^{2}W^{\prime}(\varphi_{0})}{h_0} \
\end{align}
which can be used to fix the radius of AdS$_2$,  $L$, and of the transverse space, $h_0$, in terms of our matter fields.
 \paragraph{} If we set $k=0$ in these expressions then our solutions reduce to the familiar situation of a translationally invariant AdS$_2\times R^{d}$ supported by an electric flux. Additionally, we can obtain translationally invariant fixed points at $k \neq 0$ provided that $W(\varphi_0) = 0$. In this case the lattice corresponds to an irrelevant operator and hence translational symmetry is restored in the infra-red. However by setting $\rho =0$ we can also obtain neutral AdS$_2 \times R^{d}$ geometries in which it is the lattice fields supporting the extremal horizon \cite{andrade, donos1, donos2, blaise2}. For a general solution, we have both the Maxwell field and Q-lattice present and so our infra-red fixed point has both a finite density and some non-trivial momentum relaxation.
\paragraph{} Our interest then is in the studying geometries \eqn{eq:Qansatz} with a flow to one of these fixed points in the infra-red. At a small finite temperature, the near horizon limit of such a geometry will then be described by a small black hole
\begin{align}\label{eq:small_bh_horizon}
f =L\,((r-r_{0})^{2}-r_{h}^{2}),\quad h=h_{0}, \quad \varphi=\varphi_{0}\,.
\end{align}
where in this coordinate system we now have an external horizon at $r=r_{0}+r_{h}$. For convenience we will chose to fix our coordinates such that the extremal horizon is located at $r_{0}=0$. We can then read off the temperature of this black hole as
\be
T = \frac{f'(r_h)}{4 \pi} = \frac{L r_h}{2 \pi} 
\ee
\subsection*{Thermoelectric Response Coefficients}
\paragraph{} The reason these Q-lattice models are so useful for studying transport is that it is possible to obtain simple analytic expressions for the 
thermoelectric response coefficients. In particular the DC electrical $(\sigma)$, thermoelectric ($\alpha$) and heat ($\bar{\kappa}$) conductivities can be expressed directly in terms of black
hole horizon data \cite{univdc, lattices, donos2, donos3, donos5, donos6}. The leading behaviour at low temperatures can then be written in terms of the infra-geometry \eqn{eq:small_bh_horizon} as
\begin{eqnarray} 
\sigma &=& h_0^{d/2-1} Z(\varphi_0)+ \frac{4 \pi \rho^2}{k^2 W(\varphi_0) s_0} \nonumber  \\
\alpha &=& \frac{4 \pi \rho}{k^2 W(\varphi_0)} \nonumber  \\
 \frac{\bar{\kappa}}{T} &=& \frac{4 \pi s_0 }{k^2 W(\varphi_0)}
 \label{thermoresponsemaintext}
\end{eqnarray} 
where $s_0 = 4 \pi h_0^{d/2}$ is the ground state entropy density. 
\paragraph{} As we discussed in the introduction, in order to calculate the thermal diffusivity of these theories we wish to make use of the Einstein relation
\be
D = \frac{\kappa}{c_\rho}
\label{einsteinrelation}
\ee
where $c_{\rho} = T (\partial s/\partial T)_{\rho}$ is the specific heat at fixed charge density. For particle-hole symmetric theories, this form of the Einstein relation is very familiar. However, for a generic finite density theory the coupling between charge and energy fluctuations results in a more complicated set of Einstein relations and hence the ratio $\kappa/c_{\rho}$ is not directly related to a diffusion constant. Nevertheless, as we show in Appendix~\ref{sec:einsteinrelations}, we find that in the infra-red limit there is a dramatic simplification in the Einstein relations for our AdS$_2 \times R^{d}$ theories. As a result we establish that even in our finite density models there is a thermal diffusion constant given by \footnote{We note that it was recently shown that this Einstein relation can be used to define the thermal diffusivity at finite density for a critical Fermi surface model \cite{patel}.}\eqn{einsteinrelation} .
\paragraph{}Indeed, the only subtlety we need to be aware of is that at finite density the Einstein relation is formulated in terms of the thermal conductivity at zero electrical current $\kappa = \bar{\kappa} - \alpha^2 T/\sigma$. Evaluating this using our expressions \eqn{thermoresponsemaintext} tells us that at low temperatures $\kappa$ is given by
\be
\frac{\kappa}{T} = \frac{4 \pi s_0  Z(\varphi_0) h_0^{d-1}}{\rho^2 + k^2 W(\varphi_0) Z(\varphi_0) h_0^{d-1}}
\label{kappa}
\ee
It is worth emphasising that the behaviour of this thermal conductivity is quite distinct from the other transport coefficients \eqn{thermoresponsemaintext}. In particular, in the translationally invariant limit $k^2 W(\varphi_0) \to 0$ the expressions in \eqn{thermoresponsemaintext} diverge and hence these conductivities can be sensitive to irrelevant deformations. Conversely, $\kappa$ remains finite in this limit due the presence of the net charge density. This thermal conductivity is then an intrinsic property of our infra-red theory, and is insensitive to the details of momentum relaxation\footnote{The fact that $\kappa$ can be insensitive to momentum relaxation at finite density has been emphasised in \cite{wflaw}.}.
\paragraph{} We can make this more explicit by clarifying the form of this thermal conductivity. Whilst \eqn{kappa} looks rather complicated, it can be simplified using the equations that govern our extremal geometry. Specifically, the AdS$_2$ radius $L$ is fixed through \eqn{eq:ext_constraints} in terms of the charge density and scalar fields. Using this, we find that the thermal conductivity can always be written as 
\be
\frac{\kappa}{T} = \frac{(4 \pi)^2 h_0^{d/2-1}}{2 L} 
\label{kapparesult}
\ee
and hence is determined entirely by geometric properties of the AdS$_2 \times R^{d}$ horizon. The only way the charge density $\rho$ and lattice fields $k$ enter is encoded in their effects on this background geometry.
\section{Diffusion and the Butterfly Velocity}
\label{sec:diffusion}
\paragraph{} We now have everything we need to address our main question of interest, which is to establish the relationship \eqn{butterflyresult} between this thermal diffusivity and the butterfly velocity. 
However, unlike the majority of holographic theories studied previously \cite{butterfly, incoherent}, it is not possible to extract these quantities solely from the fixed point \eqn{eq:small_bh_horizon}.

 \paragraph{} For the case of the diffusion constant, it is simple to see why we have a problem. Indeed, whilst the thermal conductivity \eqn{kapparesult} can be determined from AdS$_2\times R^{d}$ we cannot yet calculate the specific heat because the entropy density $s_0 = 4 \pi h_0^{d/2}$ is a constant.  
 \paragraph{} Crucially, an analogous pathology appears if one attempts to calculate the butterfly velocity directly in our AdS$_2 \times R^{d}$ solutions \eqn{eq:small_bh_horizon}. To see why, we can recall that for theories dual to classical gravity, the chaos parameters can be extracted by studying the construction of a shock wave geometry on the black hole horizon \cite{butterflyeffect, stringyeffects,localised}. For a general metric of the form \eqn{eq:Qansatz} one finds that chaos is described by a maximal Lyapunov exponent $\lambda_L = 2 \pi T$ and a butterfly velocity \cite{butterfly, roberts}  
\be
v_B^2 = \frac{4 \pi T}{d h'(r_h)}
\label{butterflyvelocity}
\ee
which is ill-defined when $h = h_0$ is a constant.
%
%
\paragraph{} In order to evaluate these quantities, we therefore need to consider adding irrelevant deformations to \eqn{eq:small_bh_horizon} which describe the flow of our geometry towards the infra-red fixed point. In Appendix~\ref{sec:domainwall} we will explicitly construct domain wall expansions that interpolate between our AdS$_2\times R^{d}$ solutions and the UV. Whilst the full details of these solutions are rather complicated, all that we need to calculate $c_{\rho}$ and $v_B$ are the leading corrections to $h(r)$ in the near-extremal limit. 
\paragraph{} For generic domain wall solutions these will simply take the form 
\be
h(r) = h_0 + c_h^{0}(\rho) r + \dots 
\label{metricexpan}
\ee
where $c_h^{0}(\rho)$ is a constant that is fixed by the UV data of the domain wall solution. From the point of view of our infra-red fixed point, this expansion \eqn{metricexpan} corresponds to turning on a source $c_{h}^{0}$ for a universal dilaton operator with scaling dimension $\Delta = 2$. Additionally there is a second irrelevant mode corresponding to perturbations of the scalar field $\varphi$. However, provided that this scalar mode has an IR scaling dimension $\Delta_{\varphi} > 3/2$, then we find that \eqn{metricexpan} will give the leading behaviour in $h(r)$ in the low temperature limit. 
\paragraph{} In this generic situation it is then a straightforward matter to calculate both the specific heat and the butterfly velocity from this expansion. Indeed, the key point is that it is precisely this same deformation that determines both the diffusion constant and $v_B$. As such  we will be able to obtain a simple relationship between them.
\paragraph{} Firstly, we can look at the butterfly velocity. From our metric \eqn{metricexpan} we can deduce that we now have a finite butterfly velocity given by 
\be
v_B^2 = \frac{4 \pi T}{d c_h^{0}}
\label{vbsquare}
\ee
It is interesting to note that this scaling is not what one would naively have expected in a locally critical theory (i.e. $v \sim T$) \cite{butterfly,roberts}. This reflects the fact that we needed to turn on
a dangerously irrelevant deformation to define $v_B$. The butterfly velocity therefore shows an enhanced scaling $v_B \sim \sqrt{T}$ at low temperatures. 
\paragraph{} Likewise, it is also easy for us to now determine the diffusion constant. Evaluating \eqn{metricexpan} on the horizon allows us to compute that the entropy density is given by
\be
s = 4 \pi h(r_h)^{d/2} = 4 \pi h_0^{d/2} + \frac{4 \pi^2 d }{L} h_0^{d/2-1} c_h^{0}(\rho) T + \dots 
\label{specificheat}
\ee
and hence deduce that we have a linear specific heat $c_{\rho} \sim T$ with coefficient
\be
\bigg(\frac{\partial s}{\partial T}\bigg)_{\rho} = \frac{4 \pi^2 d}{ L} h_0^{d/2 - 1} c_h^{0}  
\label{coefficient} 
\ee
Together with our expression for the thermal conductivity \eqn{kapparesult} we can use this in the Einstein relation to deduce that the diffusion constant is given by
\be
D = \frac{2}{d c_h^{0}} = \frac{v_B^2}{2 \pi T}
\ee
As such we precisely have a relationship of the form \eqn{butterflyresult} with a universal coefficient $E=1$. In particular, it is now clear that this relationship is independent of both the charge density $\rho$ and the lattice sources $k$. Indeed, provided \eqn{metricexpan} captures the leading infra-red behaviour in $h(r)$ then we obtain this same result for any of our AdS$_2 \times R^{d}$ fixed points. 
 \paragraph{} As we suggested above, a more complicated analysis is required in the special case that the scalar mode has dimension $1 < \Delta_{\varphi} < 3/2$. In such a scenario, the gravitational back-reaction of this mode becomes important in the infra-red and hence modifies the form of $h(r)$. In Appendix~\ref{sec:domainwall} we present a detailed treatment of this situation.  We now find modified scalings in the diffusion constant $D \sim T^{3 - 2 \Delta_{\varphi}}$ and butterfly velocity $v_B \sim T^{2 - \Delta_{\varphi}}$. Nevertheless, these quantities continue to obey a simple relationship of the form \eqn{butterflyresult}, but now with a different constant of proportionality $E$ that is fixed by the scaling dimension $\Delta_{\varphi}$. Whilst we are not able to obtain a closed form expression for $E$, it is simple to establish that it always lies in the range $1/2 < E \leq 1$.  
\section{Discussion} 
\label{discussion}

\paragraph{} In this paper we have calculated both the thermal diffusivity \eqn{thermaldiffusivity} and the butterfly velocity for a general family of holographic models that flow to AdS$_2 \times R^{d}$ fixed points in the infra-red. We found that both of these quantities were determined by the same irrelevant deformation of AdS$_2$ and hence established the simple relationship \eqn{butterflyresult} between them. In particular, when this deformation corresponded to the $\Delta = 2$ mode that parameterises the horizon volume we always found  $D  = v_B^2/(2 \pi T)$. 
\paragraph{} It is interesting to compare what we have seen with previous results in the literature. Specifically, in \cite{incoherent} one of us studied diffusion in certain neutral black hole geometries with broken translational symmetry. We found that when momentum relaxation was a strong effect then the diffusion constant $D = \kappa/c_{\rho}$ and butterfly velocity of these theories were related. In particular for a specific linear axion model we obtained \eqn{butterflyresult} with a coefficient $E=1$. With the benefit of hindsight, we can now understand that the reason we found this result was because in this limit the geometry sourced by the axion fields had a flow \eqn{metricexpan} towards an AdS$_2 \times R^{2}$ fixed point.
 \paragraph{} However we have shown in this paper that this connection between diffusion and chaos holds for far more general AdS$_2 \times R^{d}$  geometries. In particular we found that the relationship \eqn{butterflyresult} also applies to the diffusion constant of our finite density models.  As such, it is not necessary to consider theories with strong momentum relaxation to obtain \eqn{butterflyresult}.  Indeed we have seen that this result also applies to translationally invariant electric-AdS$_2$ and for irrelevant Q-lattices that flow to such a fixed point\footnote{In clean systems it is possible to identify an `incoherent' diffusion constant that is insensitive to momentum relaxation \cite{richardblaisesean}. It would be interesting to understand the connection between this diffusion constant and the butterfly velocity more generally.}.
%
%
%
%
%
\paragraph{} Whilst we have focused on the thermal diffusivity, one can also define a charge diffusion constant for these models via the Einstein relation $D_{c} = \sigma/\chi$ where $\chi$ is the charge susceptibility (see Appendix~\ref{sec:einsteinrelations}). However, since the electrical conductivity \eqn{thermoresponsemaintext} diverges in the translationally invariant limit then $D_{c}$ is sensitive to the strength of momentum relaxation. Moreover, in our AdS$_2$ theories the susceptibility $\chi$ is not an infra-red quantity but rather depends on the full details of the geometry \cite{iqballiu, kss1}. As such one would not expect $D_{c}$ to display a sharp connection to chaos in these general AdS$_2$ solutions\footnote{In \cite{incoherent} we found that the charge diffusion constant of the neutral axion model also obeyed a simple relationship with the butterfly velocity $D_{c} = v_B^2/(\pi T)$. However as we have remarked above we do not expect $D_{c}$ to show a sharp connection to chaos in more general AdS$_2$ geometries.}. 
\paragraph{} Finally, we note that the connection between diffusion and chaos has recently been studied in the context of a coupled Sachev-Ye-Kitaev model in \cite{stanford}. Whilst their model did not have a global charge, they were able to calculate both the diffusion constant $D = \kappa/c_{\rho}$ and the butterfly velocity. Since these SYK models do not contain any operators with $\Delta < 3/2$, we can compare their results with our findings when the leading deformation of AdS$_2 \times R^{d}$ is the dilaton \cite{yang}. Interestingly, both the diffusion constant $D \sim T^{{0}}$ and the butterfly velocity $v_B \sim {\sqrt T}$ of this SYK model show exactly the same scalings that we found for such flows. Moreover the relationship between them is also given by the formula \eqn{butterflyresult}, with precisely the same coefficient $E=1$. 
\paragraph{Note added:} Whilst this work was in preparation we were made aware of \cite{Davison:2016ngz} which also studies diffusion in AdS$_2$ geometries. 
\acknowledgments{MB acknowledges useful conversations with Richard Davison and Subir Sachdev. We are grateful to Sean Hartnoll for valuable comments on a draft of this paper. This work is supported by the Office of High Energy Physics of U.S. Department of Energy under grant Contract Number  DE-SC0012567.}

 \appendix
 
 \section{Einstein Relations near AdS$_2$}

\label{sec:einsteinrelations}
\paragraph{} In this appendix we wish to justify that in our finite density AdS$_2$ geometries there is always a `thermal' diffusion constant given by the Einstein relation $D =\kappa/c_{\rho}$. In particular, we emphasise that for generic finite density systems then this simple form of the Einstein relation is certainly not valid. Rather, fluctuations in the charge $\delta \rho$ and energy $\delta \epsilon$ densities are described by a pair of coupled diffusion equations 
\[
 \Bigg( \;\; \begin{matrix} \partial_t \delta \rho \\ \partial_t  \delta \epsilon \end{matrix} \;\; \Bigg) = {\cal D} \Bigg( \;\; \begin{matrix} \nabla^2 \delta \rho \\ \nabla^2 \delta \epsilon \end{matrix} \;\; \Bigg) 
\]
%
%
%
%
%
%
%
\paragraph{} These diffusion equations can be decoupled in terms of eigenmodes of ${\cal D}$, which describe two linear combinations of the charge and energy densities that diffuse independently\footnote{An exception is provided by translationally invariant theories, for which there is just a single `incoherent' diffusion mode \cite{richardblaisesean}.}. The diffusion constants of these modes are then simply the corresponding eigenvalues  $D_1, D_2$. The Einstein relations then relate these diffusion constants to the thermoelectric response coefficients $\sigma, \alpha, \kappa$. For a generic finite density theory these take the form \cite{hartnoll}
\begin{eqnarray}
D_{1} D_{2} &=& \frac{\sigma}{\chi} \frac{\kappa}{c_{\rho}} \nonumber \\
D_{1} + D_{2} &=& \frac{\sigma}{\chi} + \frac{\kappa}{c_{\rho}} + \frac{T \sigma}{c_{\rho}} {(\xi /\chi -   \alpha/\sigma)^2}
\label{einstein}
\end{eqnarray}
where we have defined the thermodynamic susceptibilities
\be
\chi = \bigg( \frac{\partial \rho }{\partial \mu} \bigg)_{T}  \;\;\;\;\;\;\;\;\; \xi = \bigg( \frac{\partial s}{\partial \mu} \bigg)_{T} \;\;\;\;\;\; c_{\rho} = T  \bigg( \frac{\partial s}{\partial T} \bigg)_{\rho}
\ee
 For our holographic theories, there are well known expressions for the response coefficients and susceptibilities and so \eqn{einstein} provides the most direct way to calculate the diffusion constants. However, since we would like to study $D_1$ and $D_2$ in general AdS$_2 \times R^{d}$ geometries, we need to understand which of these quantities can be determined by infrared physics. As we discussed in the main text, the thermoelectric conductivities can be tied to the black hole horizon data, and hence at low temperatures are indeed determined by the AdS$_2$ horizon \eqn{thermoresponsemaintext}. 
\paragraph{} The question of whether the susceptibilities are determined by infra-red physics is more subtle. In particular, in order to evaluate the thermodynamic derivatives $\chi$ and $\xi$ we would require knowledge of the chemical potential and hence the full details of the geometry \cite{kss1, iqballiu}. Whilst for some theories it is possible to argue that these susceptibilities are dominated by the infra-red region \cite{butterfly}, this is not the case for our AdS$_2$ models.
\paragraph{} More promising however is the behaviour of the specific heat. As we discussed in Section~\ref{sec:diffusion}, whilst this is not determined by the the extremal geometry it is an infra-red quantity that can be extracted from the irrelevant deformations of AdS$_2$. Similarly, whilst neither $\chi$ nor $\xi$ themselves are infra-red quantities, their ratio $\xi/\chi = (\partial s/ \partial \rho)_{T}$ is indeed related to the black hole thermodynamics. At low temperatures we can therefore extract it from the horizon of the extremal black hole. 
\paragraph{} Interestingly, for all our AdS$_2$ models we find that this quantity satisfies a constraint that relates it to the thermoelectric conductivities $\sigma$ and $\alpha$. Extracting this ratio by varying the second of equation in \eqn{eq:ext_constraints} with respect to $\rho$ we find that at low temperatures we have
\be
\bigg( \frac{\partial s} {\partial {\rho}} \bigg) = \frac{\alpha}{\sigma} + {\cal O} (T^{\beta})
\label{constraint}
\ee
with $\beta > 0$ determined by the irrelevant deformations of the fixed point. 
\paragraph{} For our purposes, the key consequence of this observation is that it will allow us to dramatically simplify the Einstein relations. In particular, we can consider the scaling of the various terms in the Einstein relations at low temperatures. Firstly let us assume that there is some non-trivial momentum relaxation at the fixed point (i.e $k^2 W(\varphi_0) \neq 0$). In that case the thermoelectric conductivities in \eqn{thermoresponsemaintext} scale as $\sigma \sim \alpha \sim \kappa/T \sim T^{0}$. Then the various terms in \eqn{einstein} are given by 
\be
\frac{\sigma}{\chi} \sim  T^{0} \;\;\;\;\; \frac{\kappa}{c_{\rho}} \sim T^{1 - \gamma} \;\;\;\;  \frac{T \sigma}{c_{\rho}} {(\xi /\chi -   \alpha/\sigma)^2} \sim T^{1 - \gamma + 2 \beta}
\label{scalings}
\ee
where $c_{\rho} \sim T^{\gamma}$. For the allowed parameter regime  $0 < \gamma \leq 1$ and $\beta > 0$ we therefore see that the cross terms in \eqn{einstein} are always subleading to $\sigma/\chi$ and $\kappa/c_{\rho}$. As such in the infra-red limit we find that there are two eigenvalues given by
\be
D_{1} =  \frac{\sigma}{\chi} \qquad\qquad D_{2} =  \frac{\kappa}{c_{\rho}}\,.
\label{decoupledeinstein}
\ee
%
 %
 %
\paragraph{}Similarly we can consider theories where the Q-lattice is an irrelevant deformation and we flow to a  translationally invariant AdS$_2$ fixed point where $k^2 W(\varphi_0) = 0$. In this case $\sigma$ and $\alpha$ will diverge at low temperatures and are
set by the dimension $\Delta_k > 1$ of the irrelevant lattice $\sigma \sim \alpha \sim T^{2- 2 \Delta_k}$. In contrast the thermal conductivity is finite at the fixed point and scales as $\kappa/T \sim T^{0}$. We therefore now have scalings 
\be
\frac{\sigma}{\chi} \sim  T^{2 - 2 \Delta_k} \;\;\;\;\; \frac{\kappa}{c_{\rho}} \sim T^{1 - \gamma} \;\;\;\;  \frac{T \sigma}{c_{\rho}} {(\xi /\chi -   \alpha/\sigma)^2} \sim T^{3 - 2 \Delta_k - \gamma + 2 \beta }
\label{scalings}
\ee
For $\Delta_k >1 $, $0 < \gamma \leq 1$ and $\beta > 0 $ these scalings again imply the diffusion constants take the form \eqn{decoupledeinstein}.
\paragraph{}Whilst we defer a more detailed investigation of diffusion in these models to future work\footnote{It would interesting to study diffusion in a hydrodynamic approach to these lattice models \cite{hydro,maghydro}.}, for now we can simply observe that in all these models the eigenvalues take the form of separate `charge' $D_c = D_1$ and `thermal' $D = D_{2}$ diffusivities. Note that whilst we have shown that the diffusion constants are given by \eqn{decoupledeinstein} in any of our models, we cannot determine the charge diffusivity (which depends on $\chi$) solely from knowledge of the infra-red geometry. Conversely the specific heat, and hence thermal diffusivity, is precisely determined by infra-red physics. As such it is possible to explicitly calculate $D = \kappa/{c_{\rho}}$ for our general AdS$_2$ geometries, which is our main goal in this paper. 
\section{AdS$_2\times R^{d}$ Domain Wall Solutions}
\label{sec:domainwall}
\paragraph{}In this appendix we wish to explain in detail how one can construct domain wall solutions that interpolate between the AdS$_2 \times R^{d}$ fixed points we introduced in Section~\ref{sec:qlattices} and the UV. As we discussed in the main text, including these irrelevant deformations is necessary in order to extract both the specific heat and the butterfly velocity. For concreteness we will present this construction for the case where there are two spatial dimensions in the boundary theory ($d=2$), although our analysis also goes through in other dimensions. 
\paragraph{} Our action is therefore 
\be
S=\int d^{4}x\,\sqrt{-g}\,\left(R-\frac{1}{2}(\partial\varphi)^{2}-V(\varphi)-\frac{1}{2}W (\varphi)\,((\partial\chi_{1})^{2} + (\partial\chi_{2})^{2})-\frac{1}{4}Z(\varphi)\,F^{2} \right) \nonumber
\label{eq:Qactionappendix}
\ee
and we are looking for solutions of the form 
\begin{eqnarray}
ds^{2}_{4} &=&-f(r)\,dt^{2}+\frac{dr^2}{f(r)} +h(r) (dx^2 + dy^2)  \nonumber \\
A=a(r) &dt& \;\;\;\;\;\; \chi_1= k x \;\;\;\;\;\;  \chi_2 = k y \;\;\;\;\;\;\;\;\; \varphi = \varphi(r) 
\label{eq:Qansappendix}
\end{eqnarray}
The equations of motions are then 
\begin{align}
h^{-1}\,\left( hf \varphi^{\prime} \right)^{\prime}-k^{2} h^{-1}\,W^{\prime}(\varphi)-V^{\prime}(\varphi)+\frac{1}{2}Z^{\prime}(\varphi)\,a^{\prime}{}^{2}&=0\notag\\
2h h^{\prime}f^{\prime}+2k^{2}h W+h^{2}\left( 2 V -f \varphi^{\prime}{}^{2}\right)+f h^{\prime}{}^{2}+h^{2}\,Z\,a^{\prime}{}^{2}&=0\notag\\
f^{\prime\prime}-k^{2}h^{-1}W-Za^{\prime}{}^{2}+\frac{1}{2}f\varphi^{\prime}{}^{2}-\frac{1}{2}fh^{-2}h^{\prime}{}^{2}&=0\notag\\
\left(h \,Z\,a^{\prime} \right)^{\prime}&=0\,.
\label{eq:eom1}
\end{align}
where we can integrate the Maxwell equation to give
\begin{align}
a^{\prime}=\frac{1}{Z h }\,\rho
\label{eq:g_first_order}
\end{align}
with $\rho$ the field theory charge density.
\paragraph{} Then these equations admit an AdS$_{2}\times {R}^{2}$ solution
\begin{align}\label{eq:Ext_Limitappendix}
f=L\,(r-r_{0})^2 \quad h=h_{0}, \quad \varphi=\varphi_{0}\,.
\end{align}
provided the constraints 
\begin{align}\label{eq:ext_constraintsappendix}
2 L&=\frac{\rho^{2}}{h_{0}^{2} Z(\varphi_{0})}+\frac{k^{2} W(\varphi_{0})}{h_{0}}\notag\\
0&=-2 V(\varphi_{0}) - \frac{\rho^{2}}{h_{0}^{2}Z(\varphi_{0})} -\frac{2 k^{2} W(\varphi_{0})}{h_0}\notag\\
0&=-2 V^{\prime}(\varphi_{0})+\frac{ \,Z^{\prime}(\varphi_{0})\ \rho^{2}}{h_{0}^{2}Z(\varphi_{0})^{2}} - \frac{2 k^{2}W^{\prime}(\varphi_{0})}{h_0} \
\end{align}
are satisfied. At a finite temperature we have 
\be
f_0=L\,((r-r_{0})^2- r_h^2) \quad h=h_{0}, \quad \varphi=\varphi_{0}\,.
\label{eq:smallbhappendix}
\ee
and as in the main text we will pick our coordinates so that $r_0=0$ to this order and hence we have $2 \pi T = L r_h$.
\paragraph{} We now want to consider adding irrelevant modes that will connect this solution back to the UV. The aim is to develop an expansion whose $T=0$ limit  will asymptote to the near horizon expansion of the domain wall solution connecting \eqref{eq:Ext_Limitappendix} and AdS$_{4}$. To achieve this, we can consider perturbing our black hole
solution as 
\begin{eqnarray}
\varphi &=&\varphi_{0}+\delta\varphi_{1}+\dots \nonumber \\
\quad f &=&f_{0}+\delta f_{1} + \dots \nonumber  \\
\quad h&=& h_{0}+\delta h_{1}+ \dots \nonumber 
\end{eqnarray}
At linearised order the equations of motion \eqref{eq:eom1} imply that these modes obey
\begin{align}
(f_{0}\,\delta\varphi_{1}^{\prime})^{\prime}-L\,\Delta_{\varphi} \left( \Delta_{\varphi} -1\right)\,\delta\varphi_{1}+\left(k^{2}h_{0}^{-2}W^{\prime}(\varphi_{0})-\rho^{2}\frac{Z^{\prime}(\varphi_{0})}{h_{0}^{3}Z^{2}(\varphi_{0})} \right)\,\delta h_{1}=&0\\
\left(\frac{\delta h_{1}}{r}\right)^{\prime}=&0\\
\delta f_{1}^{\prime\prime}+h_{0}^{-2}\,\left(\frac{\rho^{2}Z^{\prime}(\varphi_{0})}{Z^{2}(\varphi_{0})}-k^{2}h_{0}\,W^{\prime}(\varphi_{0}) \right)\delta\varphi_{1}+h_{0}^{-3}\left(2Lh_{0}^{2}+\frac{\rho^{2}}{Z(\varphi_{0})} \right)\,\delta h_{1}=&0\,.
\end{align}
where we have defined 
\begin{align}
L\,\Delta_{\varphi} (\Delta_{\varphi}-1)=k^{2}h^{-1}_{0}W^{\prime\prime}(\varphi_{0})+V^{\prime\prime}(\varphi_{0})-\frac{\rho^{2}Z^{\prime\prime}(\varphi_{0})}{2h_{0}^{2}Z^{2}(\varphi_{0})}+\frac{\rho^{2}Z^{\prime}{}^{2}(\varphi_{0})}{h_{0}^{2}Z^{3}(\varphi_{0})}\,,
\label{delta}
\end{align}
and as will shortly become clear $\Delta_{\varphi}$ corresponds to the AdS$_{2}$ dimension of an irrelevant operator in the boundary theory. Note that whilst $\Delta_{\varphi}$ is a natural quantity from the point of view of infra-red fixed point, it is clear from \eqn{delta} that it is sensitive to the full UV data of boundary theory. 
\paragraph{} The most general regular solution of the above system of equations is then 
\begin{align}
\delta h_{1}&=c_{h}(r_{h},\rho)\,r\notag\\
\delta\varphi_{1}&=c_{1}(r_{h},\rho)\,P_{\Delta_{\varphi}-1}\left(\frac{r}{r_{h}}\right)+\frac{1}{Lh_{0}^{3}\left( \Delta_{\varphi}-2\right) \left(\Delta_{\varphi}+1 \right)}\left(k^{2}h_{0}W^{\prime}(\varphi_{0}) -\frac{\rho^{2}Z^{\prime}(\varphi_{0})}{Z(\varphi_{0})^{2}}\right) c_{h}(r_{h},\rho)\,r\notag\\
\delta U_{1}&=-\frac{r_{h}^{2}}{h_{0}^{2}}\,\left(\frac{\rho^{2}Z^{\prime}(\varphi_{0})}{Z^{2}(\varphi_{0})}-k^{2}h_{0}\,W^{\prime}(\varphi_{0}) \right)c_{1}(r_{h},\rho)\int_{1}^{\frac{r}{r_{h}}}dz_{1}\,\int_{1}^{z_{1}}dz_{2}\,P_{\Delta_{\varphi}-1}(z_{2})\notag\\
&-\frac{c_{h}(r_{h},\rho)}{6\,h_{0}^{3}}\left(2Lh_{0}^{2}+\frac{\rho^{2}}{Z(\varphi_{0})}-\frac{\left(k^{2}h_{0}W^{\prime}(\varphi_{0}) -\frac{\rho^{2}Z^{\prime}(\varphi_{0})}{Z(\varphi_{0})^{2}}\right)^{2}}{Lh_{0}^{2}\left( \Delta_{\varphi}-2\right) \left(\Delta_{\varphi}+1 \right)}\right)\left( r+2\,r_{h}\right)\left(r-r_{h} \right)^{2}\,.
\label{eq:leading_pert_exp}
\end{align}
where $P_{n}(x)$ is the Legendre function. The above solution is completely determined up to the two constants of integration $c_{h}(r_h, \rho)$ and $c_{1}(r_{h}, \rho)$.  Additionally there is a third constant corresponding to a shift in the position of the extremal horizon $r_0$. We have chosen to set this constant is zero by a redefinition $r \rightarrow r - r_0$.
 \paragraph{} Now in order for the solution to approach smoothly the near horizon expansion of a domain wall in the $T \to 0$ limit we must require that these constants behave as
\be
c_{h}(r_{h}, \rho) \to c_{h}^{0}(\rho),\;\;\;\; c_{1}(r_{h}, \rho)\to c_{1}^{0}(\rho)\, r_{h}^{\Delta_{\varphi}-1}\,.
\label{eq:naive_lowT}
\ee
as $r_h \rightarrow 0$. We therefore deduce that the leading corrections to the extremal black hole are given by 

\begin{align}
\delta h_{1}&\to c_{h}^{0}(\rho)\,r\notag\\
\delta\varphi_{1}& \to  c_{1}^{0}(\rho)\,\frac{2^{1-\Delta_{\varphi}} \Gamma(2\Delta_{\varphi}-1)}{\Gamma(\Delta_{\varphi})^{2}}\,r^{\Delta_{\varphi}-1}+\frac{1}{Lh_{0}^{3}\left( \Delta_{\varphi}-2\right) \left(\Delta_{\varphi}+1 \right)}\left(k^{2}h_{0}W^{\prime}(\varphi_{0}) -\frac{\rho^{2}Z^{\prime}(\varphi_{0})}{Z(\varphi_{0})^{2}}\right) c_{h}^{0}(\rho)\,r\notag\\
\delta U_{1}&\to -\frac{c_{h}^{0}(\rho)}{6\,h_{0}^{3}}\left(2Lh_{0}^{2}+\frac{\rho^{2}}{Z(\varphi_{0})}-\frac{\left(k^{2}h_{0}W^{\prime}(\varphi_{0}) -\frac{\rho^{2}Z^{\prime}(\varphi_{0})}{Z(\varphi_{0})^{2}}\right)^{2}}{Lh_{0}^{2}\left( \Delta_{\varphi}-2\right) \left(\Delta_{\varphi}+1 \right)}\right) \,r^{3}\notag\\
&\qquad\qquad-\frac{c_{1}^{0}(\rho)}{h_{0}^{2}}\,\left(\frac{\rho^{2}Z^{\prime}(\varphi_{0})}{Z^{2}(\varphi_{0})}-k^{2}h_{0}\,W^{\prime}(\varphi_{0}) \right)\frac{2^{\Delta_{\varphi}-1}\Gamma(\Delta_{\varphi}-\frac{1}{2})}{\sqrt{\pi}\,\Gamma(\Delta_{\varphi}+2)}\,r^{\Delta_{\varphi}+1}
\label{eq:exp_expansion}
\end{align}
\paragraph{} From  \eqref{eq:exp_expansion} we can recognise that we have an irrelevant mode of dimension $\Delta_{\varphi}$ and a universal mode of dimension $\Delta =2$. The above solution corresponds to turning on sources  $c_h^{0}(\rho)$ and $c_{1}^{0}(\rho)$ for these modes. For a given solution these parameters, along with position $r_0$ of the extremal horizon, will be fixed at $T=0$ by the UV data of the domain wall. 
\paragraph{}  To see how this works explicitly, it is instructive to consider the parameter counting involved in the construction of the domain wall. If we assume our UV theory is described by an AdS$_{4}$ fixed point of unit radius, then we can can expand our functions near the boundary as 
\begin{align}
U&=r^{2}-\frac{c_{s}}{r}+\cdots \notag\\
h&=l_{h}^{2}\,r^{2}+\cdots \notag \\
a&=\mu-\frac{\rho}{l_{h}^{2}\,r}+\cdots \notag \\
\varphi&=\frac{\varphi_{s}}{r^{3-\Delta_{UV}}}+\frac{\varphi_{v}}{r^{\Delta_{UV}}}+\cdots \notag\,,
\end{align}
where we have chosen to only show the terms where free constants of integration appear. Note that there is additional constant of integration $c_{r}$ which simply shifts the coordinate $r\to r+c_{r}$ and is the part of diffeomorphisms our coordinate choice in \eqref{eq:Qansappendix} does not fix. 
\paragraph{}In total, there are therefore  six constants of integration in the UV which precisely matches the order of the system of first three equations in \eqref{eq:eom1} and equation \eqref{eq:g_first_order}. In the UV we can fix three of these constants: a choice of gauge $c_{r}=0$, the length scale $l_{h}$ and also the non-normalisable mode $\varphi_{s}$ of the scalar. This leaves us with three unfixed constants of integration in the UV. We therefore have precisely enough freedom to construct a unique solution matching onto our expansion \eqn{eq:exp_expansion} by using e.g. a double sided shooting method. The remaining constants of integration $c_{s}$, $\varphi_{v}$ and $\mu$, and the infra-red expansion parameters $r_{0}$, $c_{h}^{0}(\rho)$ and $c_{1}^{0}(\rho)$, will then be fixed at $T=0$ by the form of this solution. 
\paragraph{} Now that we have explained how to construct these solutions, we can proceed to extract the butterfly velocity and specific heat from these irrelevant modes. Indeed, whilst the form of the solution \eqref{eq:leading_pert_exp} looks complicated, all that we need are the corrections $\delta h$. These are governed by the universal $\Delta = 2$ mode and take the form $\delta h = c_h^{{0}}(\rho) r $ that we presented in the main text. In particular at this order in our expansion there is an entropy density
\be
s = 4 \pi h_0 + 4 \pi  c_h^{0}(\rho) r_h + \dots 
\ee
which gives a linear specific heat $c_{\rho} \sim T$. The thermal conductivity reads
\be
\frac{\kappa}{T} = \frac{(4 \pi)^2}{2 L}
\ee 
from which we calculate the diffusion constant $D = \kappa/c_{\rho}$
\be
D = \frac{1}{c_h^{0}}
\ee
Likewise we can extract the butterfly velocity \eqn{butterflyvelocity} from $h'(r_h)$ as
\be
v_B^2 = \frac{2 \pi T}{ c_h^{0}}
\ee
and so we have the relationship
\be
D = \frac{v_B^2}{2 \pi T}\,.
\label{resultappendix}
\ee
\paragraph{} The result \eqn{resultappendix} therefore always holds when these linearised modes in the domain wall expansion capture the leading order behaviour at low temperatures. To see when this is the case, we can examine the $T=0$ expansion of the functions $f, g, \varphi$. For a general domain wall solution, these can be expanded in a power series of the form \footnote{Note that in the case where $\Delta_{\varphi}=n+1$ or when $\Delta_{\varphi}=\frac{1}{n}+1$, for $n$ positive integer, there will be additional non-linear logarithmic terms of the radial coordinate $r$ that won't affect our argument.}
\begin{align}
f&=r^{2}\,\sum_{n_{1},n_{2}\geq0}B^{f}_{n_{1},n_{2}}\,(c_{h}^{0}(\rho))^{n_{1}}\,(c_{1}^{0}(\rho))^{n_{2}}\,r^{n_{1}+(\Delta_{\varphi}-1)\,n_{2}}\notag\\
h&=\sum_{n_{1},n_{2}\geq0}B^{h}_{n_{1},n_{2}}\,(c_{h}^{0}(\rho))^{n_{1}}\,(c_{1}^{0}(\rho))^{n_{2}}\,r^{n_{1}+(\Delta_{\varphi}-1)\,n_{2}}\notag\\
\varphi&=\sum_{n_{1},n_{2}\geq0}B^{\varphi}_{n_{1},n_{2}}\,(c_{h}^{0}(\rho))^{n_{1}}\,(c_{1}^{0}(\rho))^{n_{2}}\,r^{n_{1}+(\Delta_{\varphi}-1)\,n_{2}}\,.
\end{align}
where so far we have just been keeping the first terms in this expansion. That is at zeroth order we just had the IR geometry itself
\begin{align}
B^{f}_{0,0}=L^{2},\quad B^{h}_{0,0}=h_{0},\quad B^{\varphi}_{0,0}=\varphi_{0}
\end{align}
whilst the coefficients for $n_{1}=0$ and $n_{2}=1$ or vice versa can be read off precisely from from the linearised analysis we have just performed \eqref{eq:exp_expansion}. The higher order terms will then be fixed by solving the equations of motion order by order in our expansion. 
\paragraph{} At finite temperature, we will instead find that the solution near the $AdS_{2}$ region takes the form
\begin{align}
f&=r^{2}\left(1-\frac{r^{2}_{h}}{r^{2}} \right)\,\sum_{n{1},n_{2}\geq0}B^{f}_{n_{1},n_{2}}\,(c_{h}^{0}(\rho))^{n_{1}}\,(c_{1}^{0}(\rho))^{n_{2}}\,r^{n_{1}+(\Delta_{\varphi}-1)\,n_{2}}_{h}\,\mathcal{F}_{n_{1},n_{2}}(r/r_{h})\notag\\
h&=\sum_{n_{1},n_{2}\geq0}B^{h}_{n_{1},n_{2}}\,(c_{h}^{0}(\rho))^{n_{1}}\,(c_{1}^{0}(\rho))^{n_{2}}\,r^{n_{1}+(\Delta_{\varphi}-1)\,n_{2}}_{h}\,\mathcal{H}_{n_{1},n_{2}}(r/r_{h})\notag\\
\varphi&=\sum_{n_{1},n_{2}\geq0}B^{\varphi}_{n_{1},n_{2}}\,(c_{h}^{0}(\rho))^{n_{1}}\,(c_{1}^{0}(\rho))^{n_{2}}\,r^{n_{1}+(\Delta_{\varphi}-1)\,n_{2}}_{h}\,\Phi_{n_{1},n_{2}}(r/r_{h})\,.
\end{align}
The three dimensionless functions $\mathcal{F}_{n_{1},n_{2}}$, $\mathcal{H}_{n_{1},n_{2}}$ and $\Phi_{n_{1},n_{2}}$ admit an analytic expansion as $r/r_{h}\to 1$ while all of them approach the power law $\mathcal{H}_{n_{1},n_{2}}(y)\to y^{n_{1}+(\Delta_{\varphi}-1)\,n_{2}}$ as $y\to\infty$. Moreover, holding the temperature $2\pi T=L\,r_{h}$ fixed, demands that we impose $F_{n_{1},n_{2}}(1)=0$ for $n_{1}\neq 0$ and $n_{2}\neq 0$. In the next section we will explicitly construct $\mathcal{H}_{0,2}$.
\paragraph{} The key question we are interested then is the low temperature behaviour in the function $h(r)$. Whilst the finite temperature solution is quite complicated, the leading terms in $c_{\rho}$ and $v_B$ will arise from the modes that dominate $h(r)$ at $T=0$. It is simple to check from the above expansion that provided $\Delta_{\varphi} > 3/2$ then the leading deformation is precisely given by the linearised $\Delta= 2$ mode ($n_1 = 1$ and $n_2 = 0$) that we have studied so far. As such in this regime our linearised analysis is exact in the infra-red limit and hence \eqn{resultappendix} will always hold at low temperatures. However, we can also see that when $\Delta_{\varphi}<3/2$ then the term with $n_{1}=0$ and $n_{2}=2$ will instead dominate in the low $T$ expansion. This term arises from the gravitational back reaction of the scalar field on the metric. In the next section we will analyse the effects of this back reaction at finite temperature, and hence study how the diffusion constant and butterfly velocity are modified when this mode dominates in the infra-red.  
\section*{Backreaction of $\Delta_{\varphi} < 3/2$ modes}
\paragraph{} As we have just seen, when the dimension of the scalar is in the range $1 < \Delta_{\varphi} < 3/2$ then the back-reaction of this mode will dominate the behaviour of $\delta h$ in the infra-red limit.  In order to study this back reaction we need to carry out our domain wall expansion to second order. Doing this, we find that the at this order the correction $\delta h_{2}$ is given by  
\begin{align}
\delta h_{2}&=c_{h}^{(2)}(r_{h},\rho)\,r+\frac{r}{r_{h}}\frac{h_{0}\Delta_{\varphi}(c_{1}(r_{h},\rho))^2}{4}\,\int _{1}^{r/r_h}dy\,G_{\Delta_{\varphi}}\left(y\right)
\label{secondorderdeltar}
\end{align}
where the function $G_{\Delta_{\varphi}}(y)$ is defined as 
\begin{align}
G_{\Delta_{\varphi}}(y)&=\frac{1}{y^{2}}\frac{1}{y^{2}-1}\left((y^{2}+\Delta_{\varphi}-1)P_{\Delta_{\varphi}-1}(y)^{2}-2y \Delta_{\varphi} P_{\Delta_{\varphi}-1}(y)\,P_{\Delta_{\varphi}}(y)+\Delta_{\varphi}P_{\Delta_{\varphi}}(y)^{2} \right) \notag
\end{align}
\paragraph{} Again we need to fix the temperature dependence of $c_{h}^{(2)}(r_h, \rho)$ so that \eqn{secondorderdeltar} gives a regular solution in the limit $r_{h}\to 0$ (whilst holding $r$ fixed). There are two qualitatively different cases to consider, based on whether or not the integral in \eqn{secondorderdeltar} converges in the limit $r/r_h \to \infty$. In particular, as $y \rightarrow \infty$ the asymptotic form of the integrand is given by 
\begin{align}
G(y)\approx -\frac{2^{2\Delta_{\varphi}-4}(\Delta_{\varphi}-1) \Gamma^{2}(\Delta_{\varphi}-\frac{1}{2})}{\pi\,\Gamma^{2}(\Delta_{\varphi})}y^{2(\Delta_{\varphi}-2)}+\mathcal{O}(y^{2(\Delta_{\varphi}-3)})\,.
\end{align}
%
%
%
and hence for $\Delta_{\varphi}>3/2$ the integral will diverge. In this case, we therefore have the leading behaviour
\begin{align}
\delta h_{2}=c_{h}^{(2)}(r_{h},\rho)\,r-\frac{h_{0}\Delta_{\varphi}(c_{1}^{0}(\rho))^2}{4}\frac{2^{2\Delta_{\varphi}-4}(\Delta_{\varphi}-1) \Gamma^{2}(\Delta_{\varphi}-\frac{1}{2})}{\pi\,\Gamma^{2}(\Delta_{\varphi})}\,r^{2(\Delta_{\varphi}-1)}  + \dots
\end{align}
and so see that the integral in \eqn{secondorderdeltar} results in a regular correction to the metric. We therefore need only require that $c^{(2)}_{h}(r_{h},\rho)$ remains finite as $r_h \rightarrow 0$, and hence can reabsorb this term into our first order constant $c_{h}(r_{h}, \rho)$. In the low temperature limit, we can explicitly see that the back-reaction from the scalars will be sub-leading to this term and hence the relationship \eqn{resultappendix} remains unchanged. 
\paragraph{} However, things are different when $\Delta_{\varphi}<3/2$. In this case the integral in \eqn{secondorderdeltar} now converges and we can set
\begin{align}
\int_{1}^{\infty}dy\,G_{\Delta_{\varphi}}(y)= - C_{\Delta_{\varphi}}\,.
\end{align}
where $C_{\Delta_{\varphi}}$ is a constant that depends only on the scaling dimension $\Delta_{\varphi}$. In the limit $r_{h} \to 0 $ the metric will therefore look like
\begin{align}
\delta h_{2}=c_{h}^{(2)}(r_{h},\rho)\,r - \frac{h_{0} \Delta_{\varphi} C_{\Delta_{\varphi}}(c_{1}^{0}(\rho))^2}{4} r_{h}^{2\Delta_{\varphi}-3}r\,,
\end{align}
and hence the integral in \eqn{secondorderdeltar} has given rise to a singular term. In order to have a smooth limit, it is therefore necessary to demand that this cancels against a singular piece\footnote{Note that in additional to this singular piece we are also allowed to have a regular term in $c_{h}^{(2)}(r_{h},\rho)$. However as before this can simply be reabsorbed into $c_{h}(r_{h}, \rho)$ by a redefinition.}  in $c_{h}^{(2)}$
\begin{align}
c_{h}^{(2)}(r_{h},\rho)= \frac{h_{0}\Delta_{\varphi} C_{\Delta_{\varphi}}(c_{1}^{0}(\rho))^2}{4} r_{h}^{2\Delta_{\varphi}-3}\,.
\end{align}
and so as $r_h \rightarrow 0$ the second order metric is now given by 
\begin{align}
\delta h_{2}= \frac{h_{0}\Delta_{\varphi} C_{\Delta_{\varphi}} (c_{1}^{0}(\rho))^2}{4}r_{h}^{2\Delta_{\varphi}-3}r + \frac{h_{0}\Delta_{\varphi}(c_{1}^{0}(\rho))^2}{4} r_{h}^{2\Delta_{\varphi}-3}r \,\int _{1}^{r/r_h}dy\,G_{\Delta_{\varphi}}\left(y\right)
\end{align}
\paragraph{} The key point is that since we have $\Delta_{\varphi} < 3/2$ this piece will dominate over the first order term \eqn{eq:exp_expansion} as $r_h \rightarrow 0$. As a result, both the behaviour of the diffusion constant and the butterfly velocity will be modified by the presence of such operators in the theory. In particular, the entropy density $s = 4 \pi h(r_h)$ is now given by
\be
s =  4 \pi h_0 +  \pi {h_{0}\Delta_{\varphi} C_{\Delta_{\varphi}} (c_{1}^{0}(\rho))^2} r_{h}^{2\Delta_{\varphi}-2} + \dots 
\ee
from which we see that the specific heat scales with a power $c_{\rho} \sim T^{2 \Delta_{\varphi}-2}$ that depends on the dimension of this 
irrelevant deformation.
\paragraph{} Similarly the diffusion constant $D = \kappa/c_{\rho}$ has a temperature dependence $D \sim T^{3 - 2 \Delta_{\varphi}}$ and
is given by  
\be
D =  \frac{2}{h_0 \Delta_{\varphi} C_{\Delta_{\varphi}} (\Delta_{\varphi} - 1 )( c_{1}^{0}(\rho) )^2  }r_h^{3 - 2 \Delta_{\varphi}}  
\label{diffdelta}
\ee
Finally we can again extract the butterfly velocity \eqn{butterflyvelocity} from\footnote{Here we have used that $G(1) = 1 - \Delta_{\varphi}$.} $h'(r_h)$
\be
v_B^2 = \frac{8 \pi T}{h_{0}\Delta_{\varphi}(C_{\Delta_{\varphi}}+ 1 - \Delta_{\varphi} )( c_{1}^{0 }(\rho))^2 } r_h^{3 - 2 \Delta_{\varphi}}
\label{butterflydelta}
\ee
which implies the scaling $v_B \sim T^{2 -  \Delta_{\varphi}}$. 
\paragraph{} From \eqn{diffdelta} and \eqn{butterflydelta} we see that the diffusion constant and butterfly velocity are now related by 
\be
D = \frac{(C_{\Delta_{\varphi}}+ 1 - \Delta_{\varphi} )}{2 C_{\Delta_{\varphi}} (\Delta_{\varphi} - 1)} \frac{v_B^2}{2 \pi T}
\label{deltaresult}
\ee
which again is insensitive to the details of the infra-red fixed point or the matter supporting the geometry. Whilst we do not have a closed form expression for $C_{\Delta_{\varphi}}$ it is a simple matter to check numerically that the constant of proportionality, $E$, in \eqn{deltaresult} lies in the range $1/2 < E \leq 1$.
\paragraph{} In particular upon taking the limit $\Delta_{\varphi} \to 3/2$ the constant $C_{\Delta_{\varphi}}$ diverges and we recover \eqn{resultappendix}. In contrast as $\Delta_{\varphi} \to 1$ then we have the expected scalings of a locally critical theory $D \sim T$ and $v_B \sim T$ and find that the constant of proportionality $E \to 1/2$. The existence of such bounds has highly non-trivial consequences, since we have seen that the IR dimension $\Delta_{\varphi}$ can easily be changed by tuning the UV data of the boundary theory. We therefore see that whilst this can dramatically effect the values of $D$ and $v_B$, they will always be related by \eqn{butterflyresult} with an order one coefficient.

 \end{document}